\documentclass[letterpaper, 10 pt, conference]{ieeeconf}  
\usepackage{amsmath}
\usepackage{amssymb}
\usepackage{graphicx}
\usepackage{multicol}
\graphicspath{ {./images/} }

\IEEEoverridecommandlockouts                              

\title{\LARGE \bf
Cross-Subject Transfer Learning Improves the Practicality of Real-World Applications of Brain-Computer Interfaces
}

\author{Kuan-Jung Chiang, Chun-Shu Wei, {\it Member, IEEE}, Masaki Nakanishi, {\it Member, IEEE}, \\
and Tzyy-Ping Jung, {\it Fellow, IEEE}
\thanks{Research supported in part by Army Research Laboratory (W911NF-10-2-0022) and a contract from Oculus VR, LLC.}
\thanks{K.-J. Chiang, C.-S. Wei, M. Nakanishi, and T.-P. Jung* are with Swartz Center for Computational Neuroscience, Institute of Neural Computation, University of California, San Diego, La Jolla, CA 92093, USA (phone: 858-822-7555; fax: 858-822-7556; e-mail: {\tt\small jung@sccn.ucsd.edu}
}%
}

\begin{document}

\maketitle
\thispagestyle{empty}
\pagestyle{empty}

\begin{abstract}
Steady-state visual evoked potential (SSVEP)-based brain-computer interfaces (BCIs) have shown its robustness in facilitating high-efficiency communication. State-of-the-art training-based SSVEP decoding methods such as extended Canonical Correlation Analysis (CCA) and Task-Related Component Analysis (TRCA) are the major players that elevate the efficiency of the SSVEP-based BCIs through a calibration process. However, due to notable human variability across individuals and within individuals over time, calibration (training) data collection is non-negligible and often laborious and time-consuming, deteriorating the practicality of SSVEP BCIs in a real-world context. This study aims to develop a cross-subject transferring approach to reduce the need for collecting training data from a test user with a newly proposed least-squares transformation (LST) method. Study results show the capability of the LST in reducing the number of training templates required for a 40-class SSVEP BCI. The LST method may lead to numerous real-world applications using near-zero-training/plug-and-play high-speed SSVEP BCIs.
\end{abstract}

\section{Introduction}

Brain-computer interfaces (BCIs) allow users to translate their intention into commands to control external devices, enabling an intuitive interface for disabled and/or non-disabled users \cite{c1}. Among various neuromonitoring modalities, electroencephalogram (EEG) is one of the most popular ones used for developing real-world BCI applications due to its non-invasiveness, low cost, and high temporal resolution \cite{c1}. In recent studies, steady-state visual evoked potentials (SSVEP), an intrinsic neural electrophysiological response to repetitive visual stimulation, has attracted increasing attention as a pivot in implementing BCI systems because of its characteristic of robustness \cite{c2}. With recent advances in system design and signal processing, the performance of SSVEP-based BCIs has dramatically improved in the past decade \cite{c3}. Numbers of studies have reported a variety of BCI applications including text speller \cite{c3, c4}, phone-dialing system \cite{c5}, game controller \cite{c6}, etc.

To develop a real-world SSVEP-based BCI, a sophisticated algorithm to effectively decode SSVEPs plays an important role \cite{c7}. The target identification process can traditionally be divided into two parts: 1) spatial filtering, and 2) model fitting \cite{c3, c7}. Spatial filtering techniques, which include minimum energy combination (MEC) \cite{c8}, canonical correlation analysis (CCA) \cite{c9}, and task-related component analysis (TRCA) \cite{c4}, have been introduced to enhance the signal-to-noise ratio (SNR) of SSVEPs by reducing the interference from the spontaneous EEG activities. After spatial filtering, target stimuli are identified by fitting the models of SSVEPs. Computer-generated SSVEP models consisting of sinusoidal signals have been widely used to detect SSVEPs without requiring any calibration data \cite{c8, c9}. In recent studies, it has been proven that individualized templates obtained from a calibration procedure could better characterize user-specific SSVEPs than the computer-generated models, leading to drastically improved classification performance \cite{c4, c7}.

In practice, however, because of large human variability both across individuals and within individuals over time, we need to collect calibration (training) data from each individual before each session. In general, a large amount of calibration data is required for both deriving spatial filters and templates. So, the calibration procedure is often laborious and time-consuming, hindering the practicality of BCIs in a real-world context. Many researchers have attempted to adopt transfer-learning techniques to shorten the calibration process without compromising classification accuracy \cite{c10, c11}. For instance, Yuan et al. \cite{c10, c11} proposed subject-to-subject transfer learning methods, which transfer SSVEP data from existing subjects to new ones. More recently, Nakanishi et al. made it possible to transfer SSVEP data across sessions even with different electrode montages \cite{c4}. Although these approaches achieved better performance than training-free algorithms, none of them has reached comparable accuracy obtained by using individualized calibration data.

This study proposes using a least-squares transformation (LST) to facilitate cross-subject transferring of SSVEP data for reducing the calibration data/time and enhancing classification accuracy for a new user. The LST method transforms the SSVEP data from existing subjects to fit the SSVEP templates of a new user based on a small number of new templates. That is, the proposed SSVEP BCI can leverage the transformed data from other subjects and a small amount of calibration data from the new user, to develop the spatial filters for TRCA or other template-matching approaches. This new approach was evaluated using a 40-class SSVEP dataset collected from eight subjects to assess its applicability in a high-speed SSVEP-based BCI speller.

\section{Methods}

\subsection{EEG Data}

The present study used the EEG data collected in our previous study \cite{c3}. Forty visual stimuli were presented on a 23.6-inch liquid-crystal display (LCD) screen with a refresh rate of 60 Hz and a resolution of 1,920 $\times$ 1,080 pixels. The stimuli were arranged in a 5 $\times$ 8 matrix as a virtual keyboard and tagged with 40 different frequencies (8.0 Hz to 15.8 Hz with an interval of 0.2 Hz) and 4 different phases (0, 0.5 $\pi$, $\pi$, and 1.5 $\pi$). The horizontal and vertical intervals between two neighboring stimuli were 5 cm and 1.5 cm, respectively. The stimulation program was developed under MATLAB (MathWorks, Inc.) using the Psychophysics Toolbox extensions \cite{c12}.

The dataset contained nine-channel (Pz, PO5, PO3, POz, PO4, PO6, O1, Oz, and O2) EEG signals collected from eight subjects in two experiments conducted on different days. Both sessions consisted of 15 blocks, in which the subjects were asked to gaze at one of the visual stimuli indicated by the stimulus program in a random order for 0.7 s. The subjects went through 40 trials corresponding to all the visual stimuli in each block. After each stimulus offset, the screen was blank for 0.5 s before the next trial began. The intervals between sessions were different across individuals.

\subsection{TRCA-based SSVEP detection}

TRCA is a data-driven method to extract task-related components efficiently by finding a linear coefficient that maximizes their reproducibility during task periods 
\cite{c4}. Spatial filtering based on TRCA has shown significantly improving the performance of training-based SSVEP detections \cite{c4}. In addition, the TRCA-based method was able to successfully combined with the filter bank analysis, which decomposed EEG signals into sub-band components so that independent information embedded in the harmonic components can be efficiently extracted \cite{c13}. 

In the procedure of the TRCA-based method with filter bank analysis, individual calibration data for the $n$-th stimulus are denoted as $x_n \in \mathbb{R}^{{N_C} \times {N_S} \times {N_T}}$, $n = 1,2,...,N_F$. Here $N_C$  is the number of channels, $N_S$  is the number of sampling points in each trial,  $N_T$ is the number of trials, and $N_F$  is the number of visual stimuli (i.e., 40 in this study). In the training phase, the calibration data are divided to  $N_K$ sub-bands by a filter bank and become $x^k_n \in \mathbb{R}^{{N_C} \times {N_S} \times {N_T}}$, $k = 1,2,...,N_K$. The $N_K$ was set to five in this study. For each sub-band, spatial filters $w^k_n \in \mathbb{R}^{N_C}$  can be obtained by maximizing $w^TSw$ with a constraint based on the variance of reconstructed signal. Here, $S$ is the sum of inter-trial covariance matrices, i.e. $S = \sum_i \sum_j C_{i, j}$, where $C_{i, j}$ is the covariance matrices between  $i$-th and $j$-th ($i \neq j$) trials of multi-channel EEG. After obtaining the spatial filters, individual templates are prepared. The calibration data for  $n$-th stimulus $x^k_n$ are first averaged across all the training trials as $\Bar{x}^k_n \in \mathbb{R}^{{N_C} \times {N_S}}$ in each sub-band. The individual templates $y^k_n \in \mathbb{R}^{N_S}$ are obtained by applying the spatial filter to $\Bar{x}^k_n$ as $y^k_n = (w^k_n)^T\Bar{x}^k_n$.

 In the testing phase, single-trial testing data $\hat{x} \in \mathbb{R}^{{N_C} \times {N_S}}$ also go through the filter bank analysis to be decomposed into  $N_K$ sub-bands. The spatial filters $w^k_n$ obtained in training phase are then applied to the testing signals $\hat{x}^k \in \mathbb{R}^{{N_C} \times {N_S}}$ in each sub-band. Feature values $\rho^k_n$ are calculated as correlation coefficients between the testing signals and the individual templates as $\rho^k_n = r((w^k_n)^T\hat{x}^k, y^k_n)$, where $r(a, b)$ indicates the Pearson’s correlation analysis between two variables $a$ and $b$. A weighted sum of the ensemble correlation coefficients corresponding to all the sub-bands was calculated as the final feature for target identification as $\rho_n = \sum_{k = 1}^{N_K} \alpha(k) \cdot \rho^k_n$. Here, the coefficient $\alpha(k)$ was defined as $\alpha(k) = k^{-1.25} + 0.25$ according to \cite{c13}. Finally, the target stimulus $\tau$ can be identified as $\tau = \underset{n}{argmax} \  \rho_n$.

\subsection{Least-Squares Transformation (LST)}

\begin{figure}[t]
      \centering
      \includegraphics[scale=0.42]{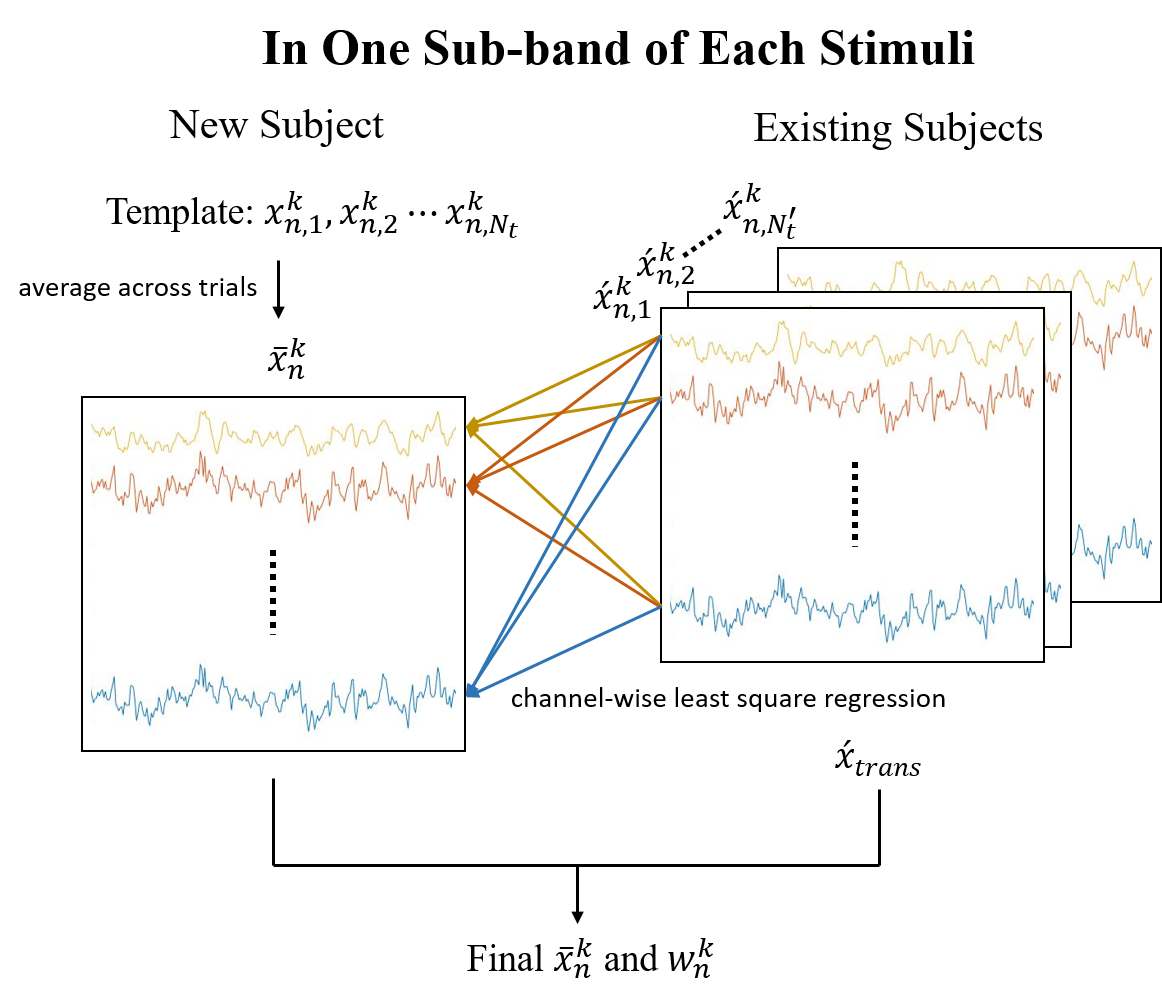}
      \caption{The procedure of transferring SSVEPs based on the least square error transformation.}
      \label{LST}
\end{figure}

Human EEG is known to present pervasive and elusive variability across individuals and even within a single subject over time\cite{c14}, posing a major obstacle in EEG data exchange across subjects. This study assumes that there exists a transformation of SSVEP signals from one subject and another. That is, if the single-trial SSVEP signals of a new user are denoted as $x$ and the ones of another existing user are denoted as $\Acute{x}$ ($x$, $\Acute{x} \in \mathbb{R}^{{N_C} \times {N_S}}$), we aim to find a transformation matrix $P \in \mathbb{R}^{{N_C} \times {N_C}}$ such that $x = P\Acute{x}$. We can acquire $P$ by applying a channel-wise least-square regression given $x$ and $\Acute{x}$, i.e. first, perform a least-square regression with $\Acute{x}$ as inputs and the first channel of $x$ as the target and second, the second channel of $x$ as the target and so on. (See Fig. \ref{LST}.)

However, to prevent the interference of noise, instead of using the single-trial signals $x$, we use the averaged signal $\Bar{x}$, obtained by averaging multiple trials of the signals from the new user. These calibration trials are called ‘template.’ Each trial of the existing users will be transformed to signals $\Acute{x}_{trans}$ which are similar to $\Bar{x}$, i.e. $\Bar{x} \approx \Acute{x}_{trans, i} = P\Acute{x}_i$ ( $i$ is the trial number). Finally, all trials of $x$ and $\Acute{x}_{trans}$ are pooled together as new training data for TRCA.

\begin{figure}[t]
      \centering
      \includegraphics[scale=0.44]{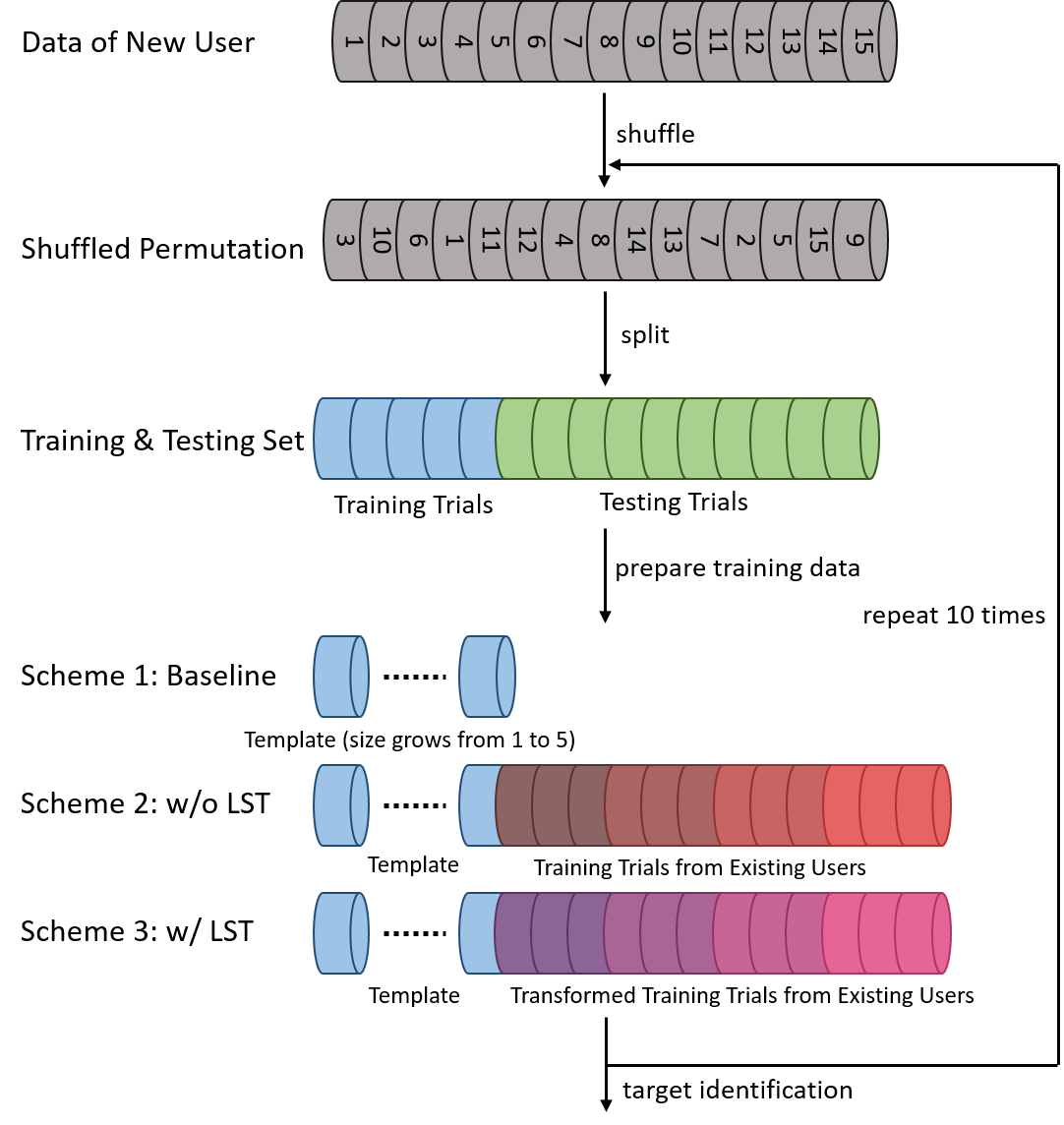}
      \caption{The flowchart of preparation of training data for three schemes.}
      \label{Flow}
\end{figure}

To validate the efficacy of the LST in transferring SSVEP data, we herein compared the SSVEP decoding performance using three schemes (shown in Fig. \ref{Flow}):

1)	Baseline: a self-decoding approach where all training data are collected from a new user (i.e., the conventional individual-template-based method).

2)	Subject-transfer without LST (w/o LST): the
training data consist of a small number of templates from the new user and a large number of trials from other subjects without any transformation.

3)	Subject-transfer with LST (w/ LST): the training data consist of a small number of templates from the new user and a large number of trials from other subjects transformed using LST.

A series of experiments were conducted to validate the performance of the proposed LST approach for the cross-subject transfer of SSVEP data. The simulation experiments focused on decoding performance in the context of real-world usage. Leave-one-subject-out (LOSO) cross-validation was employed, where a test subject plays a new user and the other subjects are existing users. When one session of the new user is being tested, the 15 trials for each stimulus was divided into 5 and 10 trials randomly as a template set and a test set. We then tested the decoding performance in that session using different sizes (1-5 trials) of templates from the template set and performed classification on the 10-trial test set with those three schemes. In ‘w/o LST’ and ‘w/ LST’ schemes, 1-5 trials of templates from the new user were used, concatenating with all trials from existing users without/with LST to form the training set for TRCA. Lastly, both sessions of the test subject were tested independently, and the random separation of template/test set was repeated 10 times. The decoding performance of each test subject was estimated by the average of 20 accuracies (2 session times 10 repeats).

\begin{figure*}[t]
      \centering
      \includegraphics[width=0.9\textwidth]{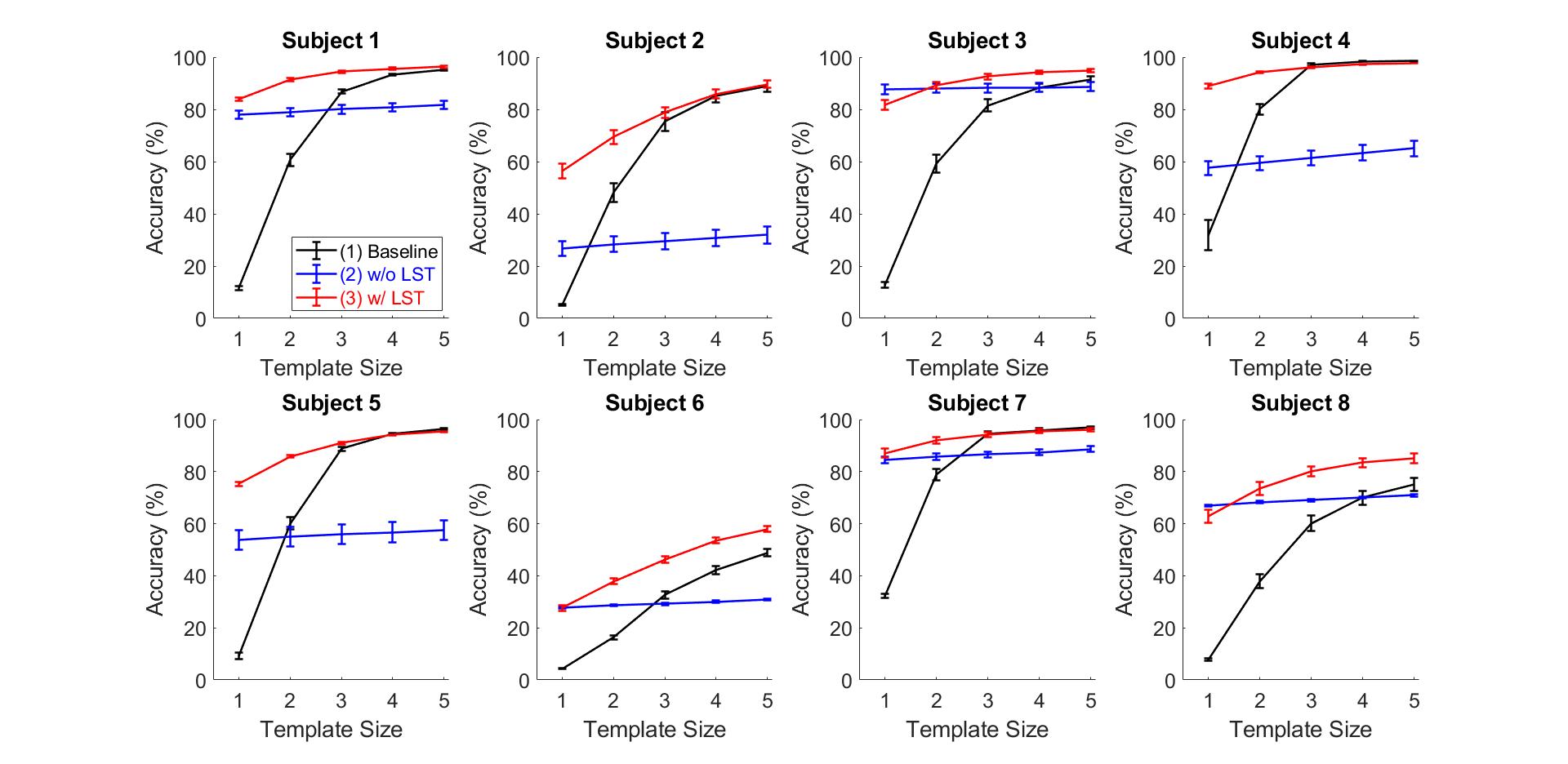}
      \caption{The averaged SSVEP decoding accuracy against template size (1-5 trials per stimulus) across 2 sessions and 10 shuffled permutations for each subject. The error-bars present the standard errors across permutations.}
      \label{result}
\end{figure*}

\section{Results}
Fig. \ref{result} compares the overall performances across all 8 subjects using those three schemes: ‘baseline’, ‘w/o LST’, and ‘w/ LST’. The result showed that ‘w/ LST’ outperformed both ‘w/o LST’ and ‘baseline’ for all subjects under most circumstances applying different template sizes. In particular, when the sizes of templates were relatively small (two or less), the LST scheme was capable of retaining the accuracy. When the template size was greater than 2, in subjects 1, 3, 6 and 8, the LST scheme achieved higher accuracy than other approaches. The LST scheme did not achieve the best accuracy only when the accuracy approximated 100\% (no room for improvement).

The overall SSVEP decoding performance is presented in Table I. A two-way repeated measures analysis of variance (ANOVA) showed significant main effects in schemes ($F(2, 119) = 12.58,\ p < 0.001$) and template sizes ($F(4, 119) = 10.32,\ p < 0.001$). The LST scheme achieved the highest overall performance regardless of the template size. In circumstances where the template size was large, ‘w/ LST’ might not have superiority against the ‘baseline’. On the other hand, when template size is as small as 1, both ‘w/ LST’ and ‘w/o LST’ outperform ‘baseline’, but no significant difference was found between these two data-transferring schemes. When template size is no less than 2, ‘w/ LST’ outperformed ‘w/o LST’. In a nutshell, the LST demonstrates its efficacy in transforming data across subjects and thus is useful for tackling insufficiency of individualized data.

\begin{table}[b!]
\begin{center}
\caption{Overall SSVEP Decoding Accuracy across Subjects Against Template Size Using LST and Other Schemes.}
\label{table_example}
\begin{tabular}{c c c c c c}
\hline
Template Size & 1 & 2 & 3 & 4 & 5\\
\hline
Baseline & 14.4\% & 55.1\% & 77.1\%	& 83.5\% &	86.4\% \\
\hline
w/o LST & 60.3\%* & 61.5\% & 62.5\%	& 63.5\%* & 64.4\%* \\
\hline
w/ LST & 70.5\%* & $\mathbf{79.2\%}$* & $\mathbf{84.2\%}$* & $\mathbf{87.4\%}$ & $\mathbf{89.1\%}$ \\
\hline
\end{tabular}
\begin{tabular}{c}
    *$p<0.05$ (vs. 'Baseline', signed rank test)\\
    \textbf{Bold}: $p<0.05$ (vs. 'w/o LST', signed rank test)\\
\end{tabular}
\end{center}
\end{table}

\section{Discussions}
The study results suggest the efficacy of the proposed LST method, which significantly enhances SSVEP decoding performance particularly when training templates are limited. While the current state-of-the-art SSVEP decoding method, template-based method with TRCA-based spatial filtering \cite{c4} (baseline), struggles with time-consuming calibration sessions, The LST is capable of leveraging existing data from other subjects and alleviating the poor decoding performance due to the lack of individual training data. As shown in Fig 3, the LST scheme presents high accuracy using a limited number of templates (down to 1 template per stimulus), and the accuracy increases with the template size.

This study validated the efficacy of the LST in transforming SSVEP templates across subjects against the pervasive human variability in the EEG data \cite{c14}. For most of the subjects, na\"ive data transferring (w/o LST) led to a lower accuracy than that of the LST, and its performance did not improve with acquiring additional templates from the user. The comparison implies that the LST is able to transform SSVEP data to model the brain response across different subjects, obviating the impact of human variability.

Finally, comparable performances were found using conventional TRCA approach (baseline) and the LST scheme when the template size grows to 5, suggesting that leveraging a large amount of data from others has no observable benefit when newly collected individual templates are sufficient. This is in line with the rationale of training-based SSVEP methods, which emphasize the importance of individualized calibration for SSVEP decoding. Nonetheless, the proposed LST method provides a satisfactory alternative source of training data and considerably reduces the calibration time for the prospective plug-and-play high-speed BCI spellers based on SSVEPs.

\section{Conclusions}
This study proposes a cross-subject transfer method, LST, for transforming SSVEP data from one subject to another. The experimental results suggest the efficacy of the LST method in alleviating the inter-subject variability in the SSVEP data and significantly improve the transferring efficiency. The improvement in the SSVEP decoding accuracy using a limited template size from a new user was very promising, suggesting a practical approach towards an online high-speed SSVEP-based BCI system with minimal calibration effort and maximal convenience and user-friendliness. Further study will be validating the LST method on different datasets including the data recorded with different headsets.

\addtolength{\textheight}{-12cm}   





\end{document}